\documentclass[letterpaper, 10 pt, conference]{ieeeconf}
\usepackage{graphicx}
\usepackage{amsmath}
\usepackage{amssymb}
\usepackage{booktabs}
\usepackage{multirow}
\usepackage{float}
\usepackage{bm}        
\usepackage{booktabs}
\usepackage{array}
\usepackage{threeparttable}
\usepackage{hyperref}

\IEEEoverridecommandlockouts                              

\makeatletter

\makeatletter
\renewcommand{\maketag@@@}[1]{\hbox{\m@th\normalsize\normalfont#1}}%

\setbox0\hbox{$\xdef\scriptratio{\strip@pt\dimexpr
    \numexpr(\sf@size*65536)/\f@size sp}$}

\newcommand{\scriptveryshortarrow}[1][3pt]{{%
    \vcenter{\hbox{\rule[\scriptratio\dimexpr-.2pt\relax]
               {\scriptratio\dimexpr#1\relax}{\scriptratio\dimexpr.4pt\relax}}}%
   \mkern-4mu\hbox{\let\f@size\sf@size\usefont{U}{lasy}{m}{n}\symbol{41}}}}

\makeatother

\title{\LARGE \bf
RAKG:Document-level Retrieval Augmented Knowledge Graph Construction
}

\author{
Hairong Zhang$^{1,2}$,
Jiaheng Si$^{1,3}$,
Guohang Yan$^{\dagger 1}$,
Boyuan Qi$^{4}$, 
Pinlong Cai$^1$, \\
Song Mao$^1$,
Ding Wang$^1$,
Botian Shi$^1$
\thanks{$^\dagger$Corresponding author. {\tt\small yanguohang@pjlab.org.cn}}
\thanks{$^1$Shanghai Artificial Intelligence Laboratory, China. }
\thanks{$^2$School of Statistics and Data Science, Nankai University, Tianjin, China. }
\thanks{$^3$Faculty of Computing, Harbin Institute of Technology, China.}
\thanks{$^4$Computer School, Beijing Information Science and Technology University, China. }
}

\begin{document}
 
\maketitle

\begin{abstract}
With the rise of knowledge graph based retrieval-augmented generation (RAG) techniques such as GraphRAG and Pike-RAG, the role of knowledge graphs in enhancing the reasoning capabilities of large language models (LLMs) has become increasingly prominent. However, traditional Knowledge Graph Construction (KGC) methods face challenges like complex entity disambiguation, rigid schema definition, and insufficient cross-document knowledge integration. This paper focuses on the task of automatic document-level knowledge graph construction. It proposes the Document-level \textbf{R}etrieval \textbf{A}ugmented \textbf{K}nowledge \textbf{G}raph Construction (RAKG) framework. RAKG extracts pre-entities from text chunks and utilizes these pre-entities as queries for RAG, effectively addressing the issue of long-context forgetting in LLMs and reducing the complexity of Coreference Resolution. In contrast to conventional KGC methods, RAKG more effectively captures global information and the interconnections among disparate nodes, thereby enhancing the overall performance of the model. Additionally, we transfer the RAG evaluation framework to the KGC field and filter and evaluate the generated knowledge graphs, thereby avoiding incorrectly generated entities and relationships caused by hallucinations in LLMs. We further developed the MINE dataset by constructing standard knowledge graphs for each article and experimentally validated the performance of RAKG. The results show that RAKG achieves an accuracy of 95.91\% on the MINE dataset, a 6.2 \% point improvement over the current best baseline, GraphRAG (89.71\%). 
The code is available at \href{https://github.com/LMMApplication/RAKG}{https://github.com/LMMApplication/RAKG}.
\end{abstract}

\section{Introduction}
With the development of LLMs \cite{bai2023qwenvlversatilevisionlanguagemodel}, their remarkable capabilities have been increasingly evident, thereby offering novel insights for further innovation across diverse domains. However, LLMs also have limitations. For example, LLMs cannot acquire knowledge beyond their training data and often omit crucial information when processing long texts. 
RAG technology \cite{NEURIPS2020_6b493230}, which leverages vector retrieval, has, to some extent, addressed the issues of delayed knowledge training and limited context length and has played a significant role in multiple fields. As RAG technology continues to evolve, the success of GraphRAG \cite{edge2025localglobalgraphrag} and Pike-RAG \cite{wang2025pikeragspecializedknowledgerationale} has further proven the importance of knowledge graphs. Therefore, establishing a comprehensive and high-quality knowledge graph is essential.

In the field of KGC, traditional methods are no longer sufficient. Rule-based approaches   \cite{gruber1993translation} are costly, inflexible, and struggle to adapt to new domains. Machine learning methods \cite{lample2016neuralarchitecturesnamedentity} rely on complex feature engineering and large amounts of labelled data, with model performance susceptible to data quality and distribution shifts. Statistical methods \cite{etzioni2008open} have high computational complexity and are particularly sensitive to data sparsity. Although new LLM-driven methods such as SAC-KG \cite{chen2024sackgexploitinglargelanguage} and KGGen \cite{mo2025kggenextractingknowledgegraphs} are emerging, their effectiveness remains to be validated, and there is a lack of unified evaluation metrics.

This study is dedicated to the construction of document-level knowledge graphs, with the assumption that each document corresponds to an ideal knowledge graph. Based on this assumption, we have established a quantitative evaluation system. Specifically, in operationalizing the concept of “closest,” we employ a dual evaluation criterion: First, concerning the topological structure, the constructed knowledge graph must comprehensively encompass all nodes present in the ideal knowledge graph. Second, regarding the relationship networks, for each corresponding node, its associated structure must attain maximum similarity with the topological relationships of the corresponding node in the ideal knowledge graph. This dual-constraint mechanism ensures both the completeness of knowledge elements and the fidelity of semantic relationships, thereby providing a quantifiable theoretical framework for assessing the quality of knowledge graphs.
To achieve the aforementioned objectives, we propose the RAKG framework, which effectively addresses the two core issues of topological structure coverage and relationship network alignment.

 \textbf{Topological Structure Coverage:} We employ a sentence-by-sentence Named Entity Recognition (NER) approach, fully capitalizing on the robust natural language processing capabilities of LLMs. In the course of this sentence-by-sentence analysis, LLMs are capable of nearly perfectly identifying entities within the text, thereby ensuring the completeness of nodes in the knowledge graph. These identified entities, acting as pre-entities, provide a solid foundation for the subsequent construction of the relationship network. 
 
 \textbf{Relationship Network Alignment:} For the construction of relationship networks, the relationship network of each node in the ideal knowledge graph is derived from the integration of all text segments where the node appears. Therefore, we propose the following two-step strategy: (1) Corpus Retrospective Retrieval. By retroactively retrieving the text segments where identified entities appear, we integrate multi-perspective semantic information and input it into the LLM for relationship network generation. (2) Graph Structure Retrieval. To maintain consistency with the initial knowledge graph, we further retrieve relevant information about the node from the initial graph and integrate this graph information into the input of the LLM.

Through these methods, the RAKG framework achieves high similarity between the relationship networks of each node in the knowledge graph and those in the ideal knowledge graph while maintaining consistency with existing knowledge.
The main contributions of RAKG are as follows:
\begin{itemize}%
    \item RAKG provides a comprehensive end-to-end solution for constructing knowledge graphs from documents. It encompasses the entire process and enables a greater focus on contextual information than traditional fragmented frameworks.
    \item RAKG introduces a progressive knowledge extraction method that is predicated on the concept of pre-entities. These pre-entities serve as intermediate representation units, and information integration is performed based on them. This approach effectively mitigates the complexity associated with entity disambiguation and circumvents the long-distance forgetting issue inherent in LLMs.
    \item In evaluating knowledge graph quality, RAKG is the first to introduce the RAG evaluation framework into the domain of knowledge graph construction. Additionally, it develops standard knowledge graphs and corresponding evaluation methods, thereby facilitating the practical assessment of the quality of constructed knowledge graphs.
    \item 
    The proposed method shows promising performance on the MINE dataset \cite{mo2025kggenextractingknowledgegraphs}; meanwhile, the related codes have been open-sourced to benefit the community.
\end{itemize}

\section{Related Works}
\label{sec:formatting}

Traditional KGC methods rely on expert systems and rule-based pattern matching \cite{suchanek2007yago}. While these methods can ensure a certain level of knowledge accuracy, they face high labour costs and poor scalability. With the development of deep learning, end-to-end construction methods based on neural networks have significantly improved the efficiency of relation extraction. In particular, the rapid growth of LLMs has provided the technical foundation for automated and large-scale knowledge graph construction \cite{zhang2024extract,lu2025karmaleveragingmultiagentllms}.

In the field of KGC, existing research mainly falls into two categories. One focuses on extracting triples from sentences \cite{suchanek2007yago}, addressing named entity recognition (NER) and relation establishment within sentences, yet ignoring entity associations across sentences. The other targets are building knowledge graphs from document-level text \cite{zhong2023comprehensive}, tackling NER, relation extraction, and entity resolution. With the rapid development of LLMs, more studies are using them for document-level knowledge graph construction. However, due to the context limitations of LLMs, simple applications often fail to meet expectations.



\subsection{Named Entity Recognition}
NER (Named Entity Recognition) is the foundational step in KGC, aiming to identify entities with specific meanings from text, such as names of people, places, and organizations. Traditional methods mainly rely on rule-based matching and dictionary lookups, but rule formulation is complex and challenging to adapt to different domains. With the development of machine learning, supervised learning methods such as Hidden Markov Models (HMMs) and Conditional Random Fields (CRFs) have been widely applied \cite{panchendrarajan2018bidirectional}, learning entity features from annotated data. However, these methods are highly dependent on annotated data. In recent years, deep learning methods such as Recurrent Neural Networks (RNNs), Convolutional Neural Networks (CNNs) \cite{zhang2023deepkedeeplearningbased}, and their variants have shown excellent performance in entity recognition, automatically learning text features to improve recognition accuracy and robustness. Moreover, applying pre-trained language models such as BERT and XLNet has further advanced entity recognition \cite{devlin2019bert}, with their rich language knowledge and contextual understanding capabilities supporting more accurate entity identification.

\subsection{Coreference Resolution}
Coreference resolution is a key task in knowledge graph construction, aiming to identify different expressions in the text that refer to the same entity and to resolve the ambiguity of entity references \cite{miculicich2022graph}. Rule-based methods perform coreference resolution by defining lexical and syntactic rules, but these rules are complicated and exhausting and are easily affected by text variations. Statistical methods train models using corpus statistical features, but feature engineering is complex, and generalization ability is limited. Machine learning methods such as Support Vector Machines (SVMs) and decision trees judge coreference by extracting feature vectors, but feature selection and parameter optimization are challenging. Deep learning methods \cite{lee2017end,wu2017deep}, such as neural network models, automatically learn text features and enhance resolution effects by incorporating attention mechanisms. The application of pre-trained models further strengthens their semantic understanding and coreference relation capture capabilities \cite{joshi2019bert}, providing new pathways for more accurate coreference resolution.

\subsection{Relation Extraction}
Relation extraction is a core task in knowledge graph construction, aiming to identify relationships between entities in text. Rule-based methods define pattern-matching relationships, but rule construction is cumbersome and challenging to adapt to different domains and text styles. Deep learning methods \cite{zeng2014relation}, such as Convolutional Neural Networks (CNNs) \cite{nguyen2015relation,santos2015classifying}, Recurrent Neural Networks (RNNs), and their improved models, are widely used and can automatically learn text features to improve the accuracy and efficiency of relation extraction. Moreover, the application of pre-trained language models such as BERT \cite{devlin2019bert} and RoBERTa has further enhanced the performance of relation extraction, with their rich language knowledge and semantic understanding capabilities providing support for more precise capture of relationships between entities \cite{cui2020revisiting}. Meanwhile, researchers continuously explore new model architectures and training strategies to improve relation extraction's performance and generalization ability.
\subsection{Retrieval-Augmented Generation}

Due to the lengthy document context in constructing document-level knowledge graphs, directly using LLMs makes it challenging to capture detailed information \cite{yang2025graphusionragframeworkknowledge}. To address this issue, we introduce the RAG (Retrieval-Augmented Generation). RAG retrieves relevant information from external knowledge bases. It inputs it as a prompt to LLMs, enhancing the models' ability to handle knowledge-intensive tasks, such as question answering, text summarization, and content generation. RAG technology typically utilizes documents and knowledge graphs as external knowledge bases (e.g., lightRAG \cite{guo2024lightrag}, GraphRAG). Innovatively, we apply the RAG concept in reverse to the knowledge graph construction process. By retrieving relevant passages and graph information, we help LLMs more accurately generate entities and their relational networks.

\begin{figure*}[ht]
    \centering
    \includegraphics[width=0.95\textwidth]{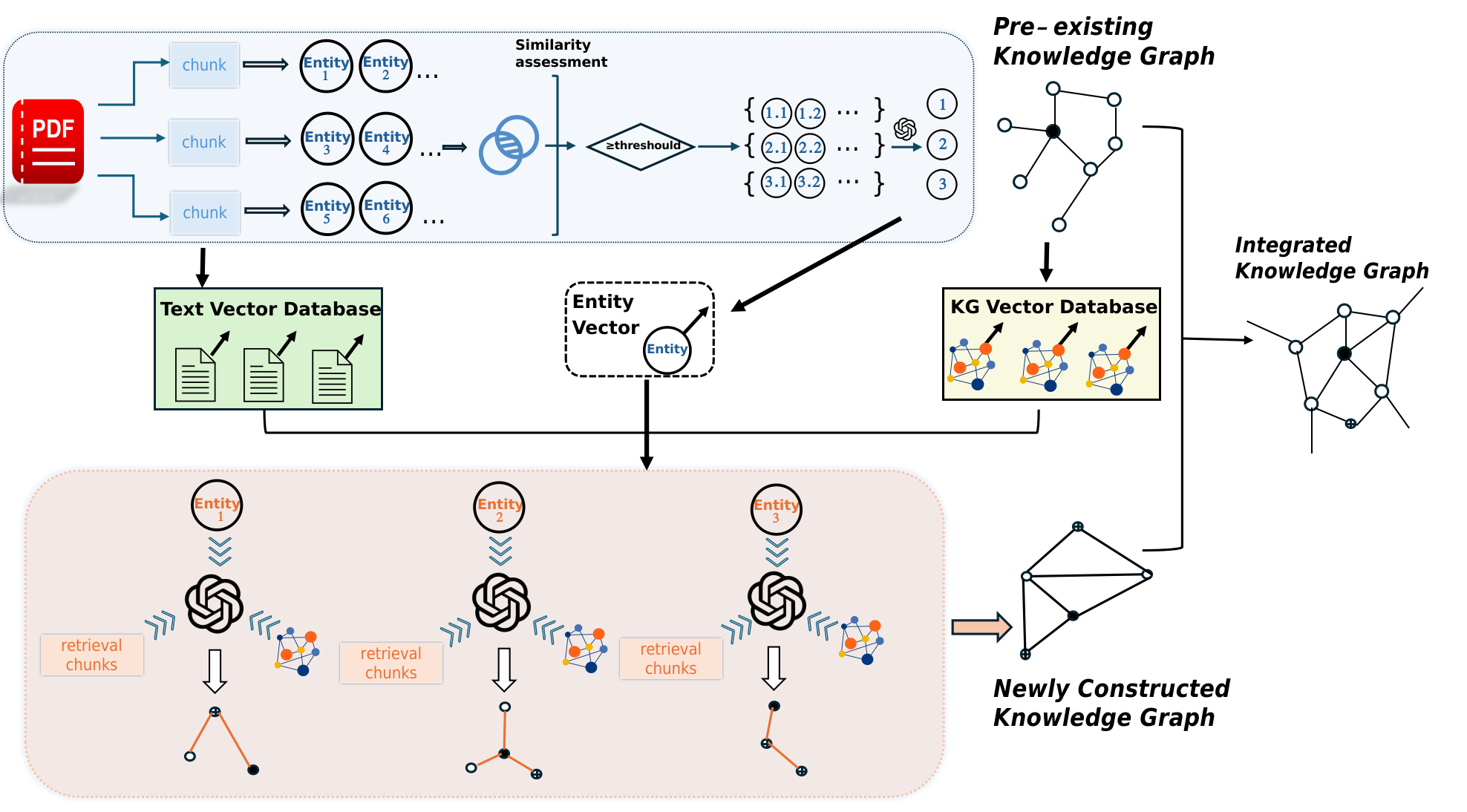} 
    \caption{The RAKG framework processes documents through sentence segmentation and vectorization, extracts preliminary entities, and performs entity disambiguation and vectorization. The processed entities undergo Corpus Retrospective Retrieval to obtain relevant texts and Graph Structure Retrieval to get related KG. Subsequently, LLM is employed to integrate the retrieved information for constructing relation networks, which are merged for each entity. Finally, the newly built knowledge graph is combined with the original one.}
    \label{fig:rakg_flow}
\end{figure*}

\section{Method}
We have developed a document-level knowledge graph construction framework, RAKG, which leverages LLMs for document-level knowledge graph construction. 
To facilitate comprehension, Table \ref{tab:notation} provides a summary of the key notations used in this paper. An overview of RAKG is shown in Figure \ref{fig:rakg_flow}.

\begin{table}[!t]
    \centering
    \normalsize
    \caption{Summary of Notation} 
    \label{tab:notation} 
    \begin{threeparttable}
    \begin{tabular}{p{1.5cm} p{6.5cm}}
        \toprule
        Notation & Description  \\
        \midrule
        $KG^*, V^*, E^*$  & Ideal knowledge graph, its set of entities, and its set of directed edges \\
        $KG', V', E'$  & Initial knowledge graph, its set of entities, and its set of directed edges \\
        $KG, V, E$  & Constructed knowledge graph, its set of entities, and its set of directed edges \\
        $D$ & Input document \\
        $T$ & Set of text chunks derived from the document \\
        $text_i$ & An individual text chunk \\
        $e$ & An entity \\
        $Pre_{entity}$ & Set of preliminary entities identified by NER from the text chunks $T$ \\
        $V_T$ & Set of vectors representing the text chunks in $T$ \\
        $V_{kg}$ & Set of vectors representing the entities in the knowledge graph \\
        $V_{Pre_{entity}}$ & Set of vectors representing the preliminary entities identified by NER \\
        $Vect(\cdot)$ & Function that vectorizes the input \\
        \bottomrule
    \end{tabular}
    \end{threeparttable}
\end{table}

\subsection{Problem Formulation} 
Given a document $D$, we assume the existence of a theoretically perfect knowledge graph construction process:
\begin{equation}
KG^* = \mathrm{Construct}(D)
\end{equation}
This ideal knowledge graph can be formally represented as:
\begin{align}
G^* &= (V^*, E^*)  \label{eq:ideal_kg} \\
KG^* &= \{ (h_i^*, r_i^*, t_i^*) \mid h_i^*, t_i^* \in V^*,\ r_i^* \in E^* \} \label{eq:kg_star}
\end{align}
Here, the set of triples $KG^*$ comprehensively covers all semantic relationships in document $D$.

The objective of this paper is to construct a  knowledge graph:
\begin{equation}
KG = \mathrm{RAKG}(D)
\end{equation}
Its formal definition is:
\begin{align}
G &= (V, E) \label{eq:real_kg} \\
KG &= \{ (h_i, r_i, t_i) \mid h_i, t_i \in V,\ r_i \in E \} \label{eq:kg}
\end{align}

The constructed knowledge graph $KG=RAKG(D)$ must satisfy the following approximation conditions:
\begin{align}
&\forall e^* \in V^* \quad \exists e \in V \quad  \\
&e^* \approx e  \\
&\mathrm{rel}(e^*) \approx \mathrm{rel}(e) \quad 
\end{align}

The relationship mapping functions are defined as:
\begin{align}
\mathrm{rel}(e^*) &= \left\{ (e^*, r_i^*, t_i^*) \mid (e^*, r_i^*, t_i^*) \in KG^* \right\} \\
\mathrm{rel}(e) &= \left\{ (e, r_i, t_i) \mid (e, r_i, t_i) \in KG \right\}
\end{align}

\subsection{Knowledge Base Vectorization}
\subsubsection{Document Chunking and Vectorization}
RAKG employs a dynamic chunking strategy based on semantic integrity rather than fixed-length divisions. Specifically, the text is segmented at sentence boundaries as shown in Equation (\ref{eq:docsplit},\ref{eq:splitedtext}), and each chunk is vectorized as shown in Equation (\ref{eq:textvector}). 

\begin{align}
T &= DocSplit(D) \label{eq:docsplit} \\
\text{s.t. } \forall i &\neq j,\, text_i \cap text_j = \emptyset \,\land\, \bigcup_{i=1}^n text_i = D  \label{eq:splitedtext}\\
V_T &= \{ \vec{v}_i \mid \vec{v}_i = Vect(text_i),\, text_i \in T \} \label{eq:textvector}
\end{align}
This approach reduces the amount of information processed by the LLM each time while ensuring the semantic integrity of each chunk, thereby improving the accuracy of named entity recognition.

\subsubsection{ Knowledge Graph Vectorization}
The initial knowledge graph is vectorized by extracting the name and type of each node and using the BGE-M3 model \cite{chen2024bgem3embeddingmultilingualmultifunctionality} for vectorization.

\begin{align}
V_{kg} &= \{ \vec{v}_j \mid \vec{v}_j = Vect(node_j),\, node_j \in KG' \}  
\end{align}

\subsection{Pre-Entity Construction}
\subsubsection{Entity Recognition and Vectorization}
Named entity recognition is performed segment by segment for the segmented text chunks. This process is completed by the LLM, which analyzes the entire text chunk to identify entities. For each pre-entity, the LLM further assigns type and description attributes. The type attribute distinguishes the category of the entity, while the description provides a brief explanation to differentiate entities with similar names. 

\begin{align}
Pre_{entity_i} &= \text{NER}(text_i) \\
Pre_{entity} &= \bigcup_{i} Pre_{entity_i} \\
\end{align}

A chunk-id attribute is added to indicate which text chunk the entity originates from. The entity’s name and type are combined and vectorized.

\begin{equation}
V_{Pre_{entity}} = \left\{ \vec{v}_i = Vect(e_i) \mid e_i \in {Pre_{entity}}\right\} 
\end{equation}

\subsubsection{Entity Disambiguation}
After completing entity recognition and vectorization for the entire document, similarity checks are performed on each entity. Entities with similarity scores above a threshold are placed into a preliminary similar entity set, which is then individually inspected by the LLM to obtain the final similar entity set. Entities in the final set are disambiguated into a single entity, with their corresponding chunk-ids linked together.

\begin{align}
Sim(e_i) &= \left\{ e_j \mid e_j \in Pre_{entity},\ VectJudge(e_i, e_j) = 1 \right\} \label{eq:sim}\\
Same(e_i) &= \left\{ e_j \mid e_j \in Sim(e_i),\ SameJudge(e_i, e_j) = 1 \right\} \label{eq:same}
\end{align}
These functions are used sequentially in the disambiguation process. First, \(\text{VectJudge}(e_i, e_j)\) is applied to efficiently filter potential matches based on vector similarity, forming a preliminary similar entity set as shown in Equation (\ref{eq:sim}). Then, \(\text{SameJudge}(e_i, e_j)\) is used to refine this set by making a final determination of identity, resulting in the final similar entity set as shown in Equation (\ref{eq:same}).

\subsection{Relationship Network Construction}
\subsubsection{Corpus Retrospective Retrieval}
For a specified entity, retrieve the associated text segments via chunk-id and use vector retrieval to obtain text segments similar to the selected entity.

\begin{align}
retriever_{V_T}(e) 
    &= \left\{ text_i \left| 
        \begin{aligned}
            &text_j \in T,\\ 
            &retriever(e, \bm{v}_j,threshold) = 1
        \end{aligned} \right. \right\}
\end{align}

\subsubsection{Graph Structure Retrieval}
Perform vector retrieval in the initial knowledge graph for a specified entity to obtain entities similar to the selected entity and extract their relationship networks.

\begin{align}
retriever_{V_{kg}}(e) 
    &= \left\{ node_j \left| 
        \begin{aligned}
            &node_j \in KG',\\
            &retriever(e, \bm{v
            }_j, threshold) = 1
        \end{aligned} \right. \right\}
\end{align}

\subsubsection{Relationship Network Generation and Evaluation}
Integrate the retrieved text and relationship networks and process this information using the LLM to obtain attributes and relationships for the central entity as shown in Equation (\ref{eq:LLMrel}). Use the LLM as a judge for the generated triplets to assess their truthfulness.
\begin{align}
rel(e_i) 
    &= LLM_{rel}\left( 
        entity_1,\,
        retriever_{text}(e_i),\,
        retriever_{kg}(e_i)
    \right) \label{eq:LLMrel}
\end{align}

\subsection{Knowledge Graph Fusion}
\subsubsection{Entity Merging}
Entities in the new knowledge graph may refer to the same entities in the initial knowledge graph. It is necessary to disambiguate and merge entities from the new knowledge graph with those in the initial knowledge graph.
\subsubsection{Relationship Integration}
To obtain a more comprehensive knowledge graph, relationships in the new knowledge graph need to be integrated with those in the initial knowledge graph.

\section{Experience}
To comprehensively evaluate the performance of RAKG across different topics and domains, we conducted experiments on the MINE dataset \cite{mo2025kggenextractingknowledgegraphs}.

\subsection{Experimental Settings}
\subsubsection{Dataset}
The MINE dataset contains 105 articles, each approximately 1000 words in length, covering multiple domains, including history, art, science, ethics, and psychology. An LLM generated these articles based on 105 diverse topics. To assess the quality of KG generation, we used the semantic inclusion rate to measure its ability to capture key information from the articles. Specifically, we extracted 15 facts for each article using the LLM and manually verified their accuracy and relevance. By checking whether the knowledge graph captured these facts, we evaluated the effectiveness of the text-to-KG extractor.

We queried the article's knowledge graph for the 15 facts for each KG generation method. We identified the top k nodes semantically closest to each point, as well as all nodes within two hops of these nodes. These nodes and their relationships were returned as query results. We then used the LLM to evaluate the results and generate binary outputs: if the fact could be inferred based solely on the queried nodes and relationships, the output was 1; otherwise, it was 0. The final MINE score for each knowledge graph generator was calculated by determining the proportion of 1s among all 15 evaluations, objectively comparing each method's ability to capture and retrieve information from the articles accurately.

\subsubsection{Baseline}
\begin{itemize}
    \item \textbf{KGGen \cite{mo2025kggenextractingknowledgegraphs}}: The Stanford Trustworthy AI Research Laboratory (STAIR Lab) developed KGGen, an open-source tool designed to generate knowledge graphs from plain text automatically. It leverages advanced language models and clustering algorithms to transform unstructured text data into structured networks of entities and relationships. Available as a Python library (installable via \texttt{pip install kg-gen}), KGGen is convenient for researchers and developers to use.
    
    \item \textbf{GraphRAG \cite{edge2025localglobalgraphrag}}: Graph-based Retrieval-Augmented Generation (GraphRAG) is a knowledge graph-based enhanced retrieval generation framework proposed by Microsoft, aiming to overcome the limitations of traditional RAG methods. Its core idea is to build a structured knowledge graph to model the global semantics of document content, thereby enhancing the performance of LLMs. The workflow of GraphRAG mainly includes three steps: graph indexing (building and indexing the knowledge graph), graph retrieval (retrieving relevant information from the knowledge graph), and graph-enhanced generation (generating text using the retrieved information). This method effectively utilizes the structural information between entities, enhancing the relevance and accuracy of the generated results.

\end{itemize}

\subsubsection{Metrics}
\begin{itemize}
    \item \textbf{Entity Density (ED)}:
    \begin{equation}
ED = N_e \label{eq:ed}
\end{equation} 

$N_e$ represents the number of entities. Generally, the more entities extracted, the stronger the ability to extract information from the text.
\item \textbf{Relationship Richness (RR)}:
\begin{equation}
RR = \frac{N_r}{N_e} \label{eq:rr}
\end{equation}
$N_r$ represents the number of relationships (attributes). Generally, the more complex the entity relationship network, the stronger the ability to extract information from the text.
\item \textbf{Entity Fidelity (EF)}:
\begin{equation}
EF = \frac{1}{N_e} \sum_{i=1}^{N_e} LLMJudge_{entity}(e_i, retriever_{V_T}(e_i)) \label{eq:ef}
\end{equation}
    LLM is used as a judge to evaluate each extracted entity and assign a value between 0 and 1, representing the entity's credibility.
    \item \textbf{Relationship Fidelity (RF)}:
    \begin{equation}
    \begin{aligned}
RF = \frac{1}{N_r} \sum_{i=1}^{N_r} LLMJudge_{rel} \\
(e_i, retriever_{V_T}(e_i), retriever_{V_{kg}}(e_i)) \label{eq:rf}
\end{aligned} 
\end{equation}
    LLM is used as a judge to evaluate each extracted relationship and assign a value between 0 and 1, representing the credibility of the relationship.
    \item  \textbf{accuracy} : 
    The accuracy of question answering on the MINE dataset is measured by the constructed knowledge graph, where higher accuracy means the graph retains the semantic info better than the original text.
    \item \textbf{Entity Coverage (EC)}:
    \begin{equation}
\text{EC} = \frac{\left| E \cap E^* \right|}{\left| E^* \right|}
\end{equation}
    The entity coverage rate reflects the proportion of entities in the evaluated knowledge graph that semantically match those in the standard knowledge graph, indicating its completeness at the entity level. It is calculated by dividing the number of entities in the intersection of the evaluated and standard knowledge graph entity sets by the number of entities in the standard knowledge graph entity set.
    \item \textbf{Relation Network Similarity (RNS)}:
    \begin{equation}
\text{RNS} = \sum_{e_i \in  E \cap E^* } \left( RelSim_i \times Entity Weight_i \right)
\end{equation}
    Relationship network similarity measures the similarity between the evaluated and standard knowledge graphs at the relationship level by calculating the similarity of the relationship networks corresponding to the same entities and combining it with entity weights. For each entity in the intersection of the evaluated and standard knowledge graphs, the relationship similarity is calculated and multiplied by the corresponding entity weight. Then, the results for all entities are summed.
    
\end{itemize}

\subsection{Experimental Results}
We conducted experiments on the MINE dataset, where the LLM we used was Qwen2.5-72B \cite{qwen2025qwen25technicalreport}, and the vector embedding model was BGE-M3 \cite{chen2024bgem3embeddingmultilingualmultifunctionality}.

\begin{figure}[htbp]
    \centering
    \includegraphics[width=0.5\textwidth, keepaspectratio]{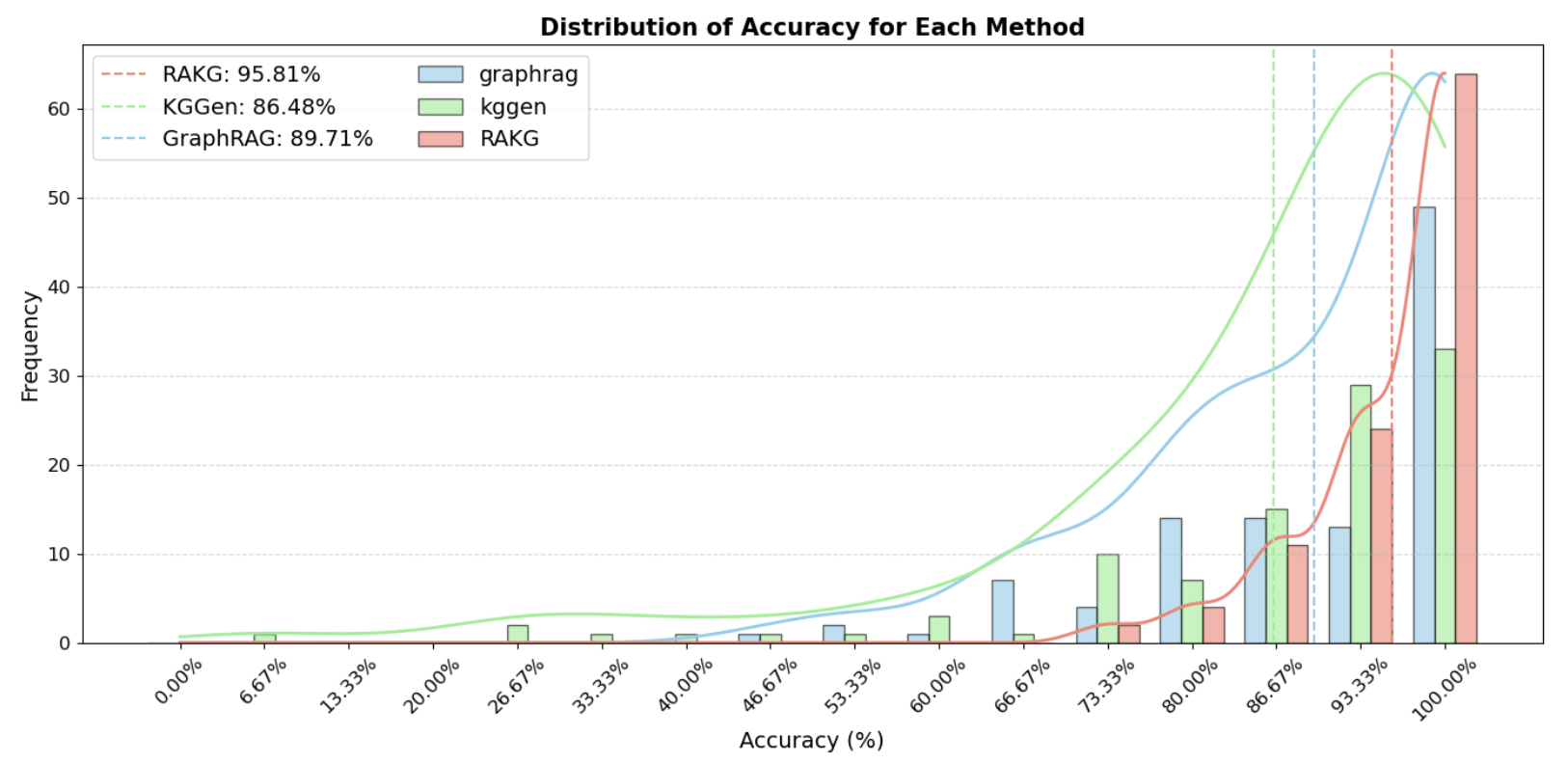}
    \caption{ Distribution of SC scores across 105 articles for GraphRAG, KGGen, and RAKG on the MINE dataset. The results demonstrate that RAKG achieves an accuracy of 95.81\%, outperforming KGGen (86.48\%) and GraphRAG (89.71\%).}
    \label{fig:accuracy_show}
\end{figure}


Compared with baseline models such as KGGen, the RAKG model has demonstrated a significant improvement in the key metric of accuracy as Figure \ref{fig:accuracy_show}, with a clear and distinct advantage. This result strongly indicates that the knowledge graph constructed by the RAKG model possesses a more substantial capability for semantic information extraction, enabling it to more comprehensively and accurately mine rich semantic knowledge from the original text. With this excellent ability in knowledge extraction and representation, the RAKG model has the potential to achieve better performance in a variety of subsequent natural language processing tasks, such as semantic search and intelligent question answering.

\begin{figure}[htbp]
    \centering
    \includegraphics[width=0.52\textwidth, keepaspectratio]{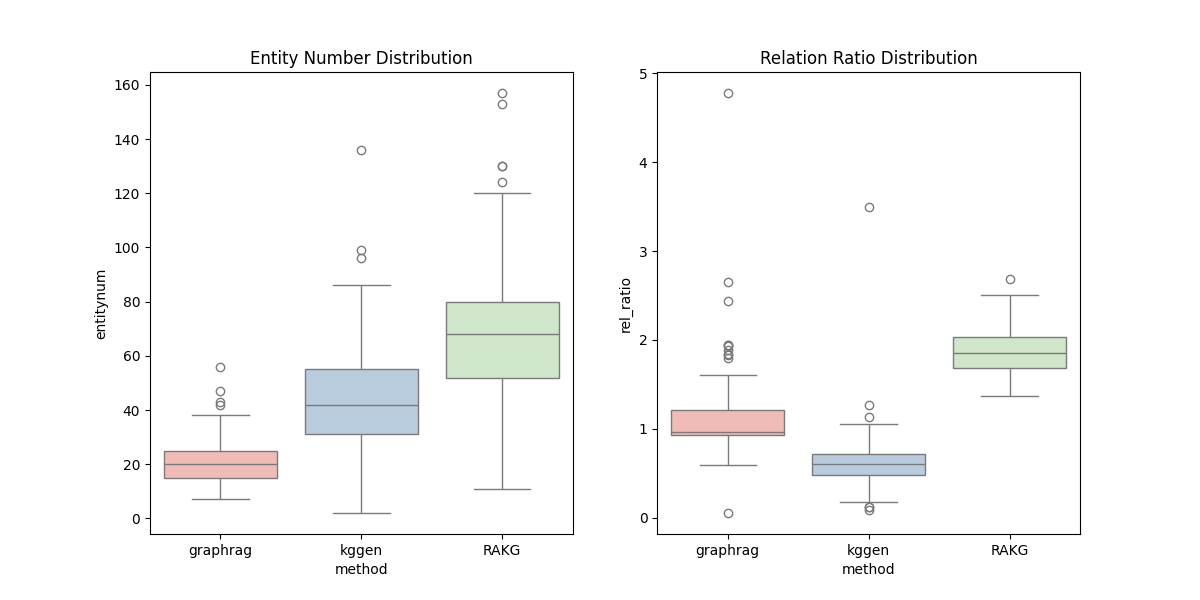}
    \caption{This visualisation of the experimental results shows the entity density and relation richness of knowledge graphs generated by RAKG, GraphRAG, and KGGen. The results indicate that RAKG produces more dense entities and richer relations than GraphRAG and KGGen.}
    \label{fig:entity_rel_show}
\end{figure}


The knowledge graph constructed by the RAKG model demonstrates a higher concentration in entity density as Figure \ref{fig:entity_rel_show}. This means that, against the same textual corpus background, RAKG can identify and incorporate a piece of richer entity information, fully excavating those entities with key semantic values hidden deep within the text. As a result, the entity composition of the knowledge graph becomes more substantial and dense.
At the same time, on the dimension of relationship density as Figure \ref{fig:entity_rel_show}, the knowledge graph built by RAKG exhibits a more complex network of associations. It can accurately capture the diverse and nuanced interaction relationships between entities and organize and present the intricate relational threads implicit in the textual context, thus weaving a dense and complex web of relationships.
These significant characteristics indicate that, in the knowledge graph construction process, the RAKG model can more efficiently mine key information from the text and more comprehensively capture entities and their relationships than other models.

\begin{figure}[ht]
    \centering
    \includegraphics[width=0.5\textwidth]{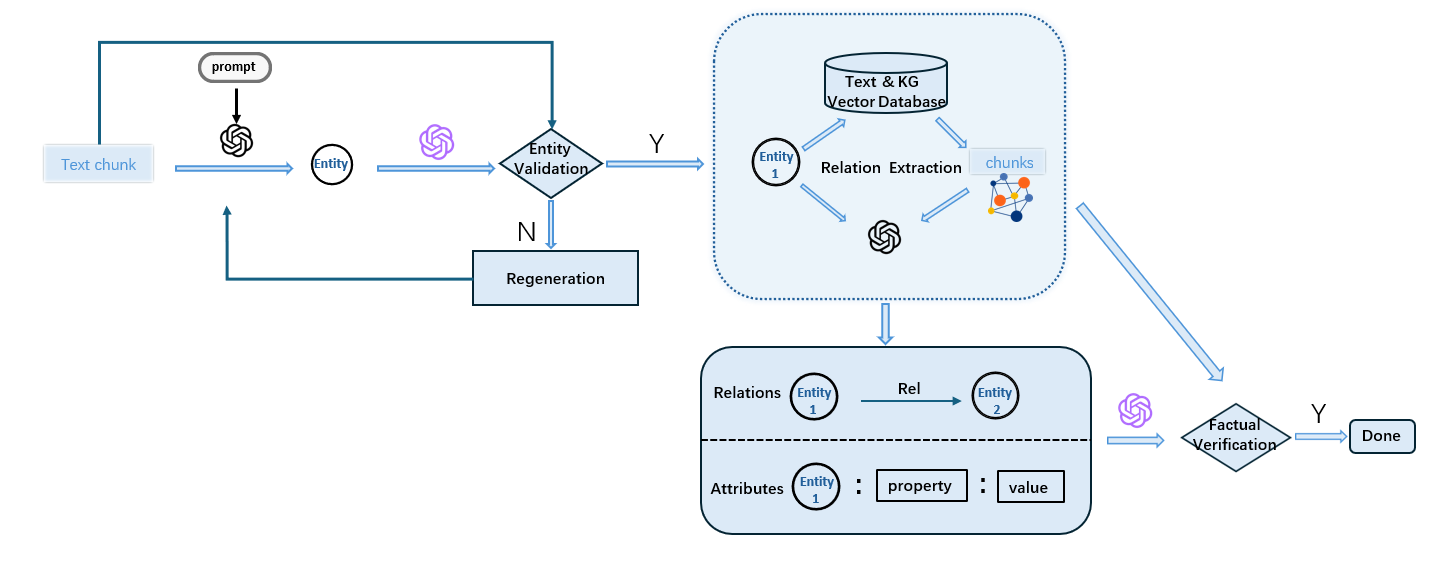} 
    \caption{The process of LLM as judge: Extracted entities are checked against the source text to eliminate hallucinations. The retriever uses entities to fetch relevant texts and KG, building a relation network. This network is then verified for consistency with the retrieved information.}
    \label{fig:llmjudge_flow}
\end{figure}
In constructing and evaluating the knowledge graph, we innovatively introduced the evaluation metric system of RAG into the KGC task to ensure the accuracy and fidelity of the constructed knowledge graph. Specifically, we leveraged the powerful capabilities of LLMs to rigorously assess the entities extracted by RAKG from the text, determining whether these entities strictly adhere to the original text's content framework and semantic logic. Meanwhile, the LLM was also employed to evaluate whether RAKG maintained high fidelity to the text when constructing the relationship network based on retriever content, ensuring that the established relationships reflect the semantic associations within the text rather than being fabricated out of thin air. Through this dual evaluation mechanism, we could precisely eliminate nodes and relationships that were falsely generated due to potential hallucinations of the LLM, thereby effectively enhancing the quality and reliability of the knowledge graph. The detailed implementation process of this approach is illustrated in  Figure \ref{fig:llmjudge_flow}.

\begin{figure}[htbp]
    \centering
    \includegraphics[width=0.51\textwidth, keepaspectratio]{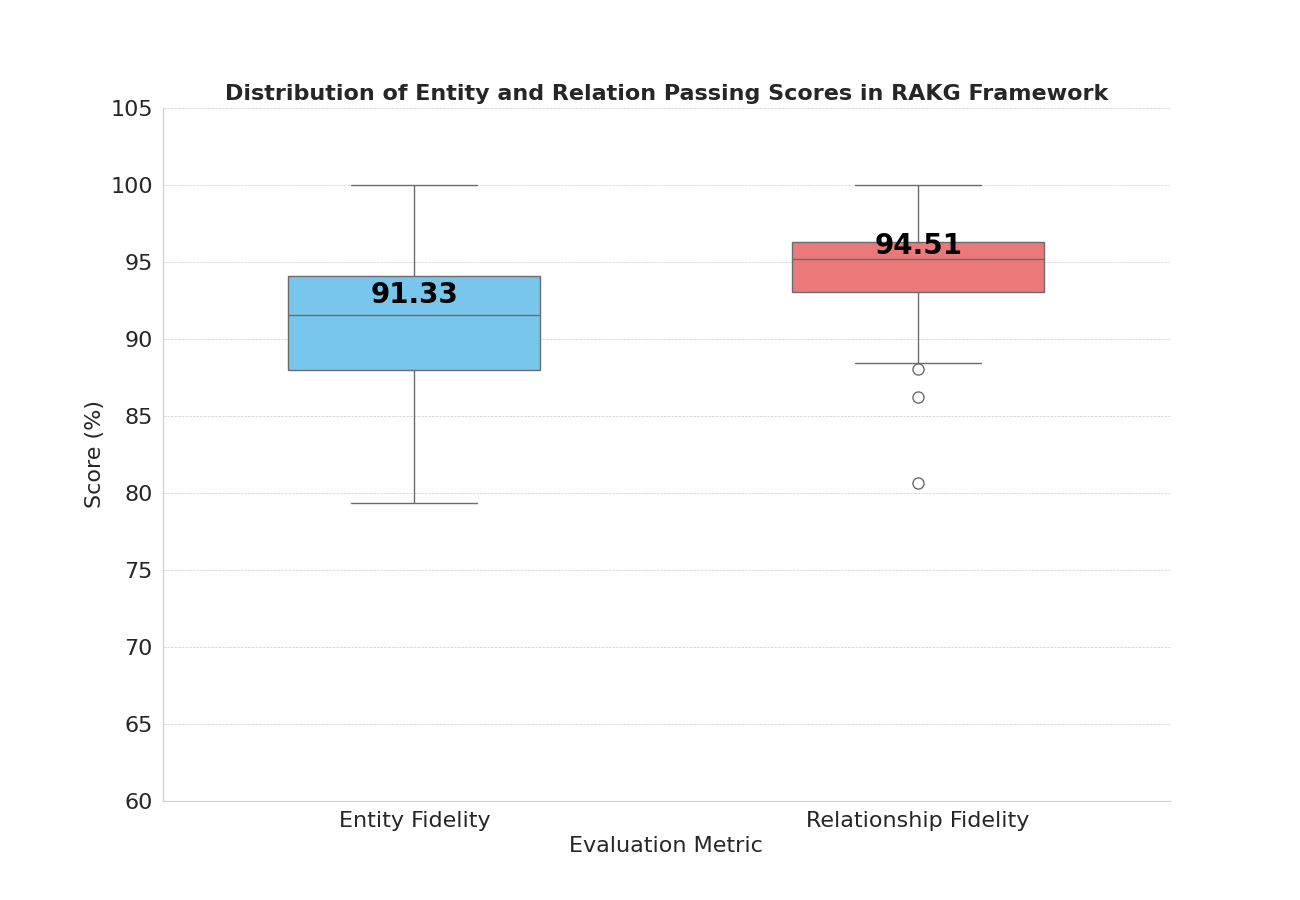}
    \caption{Results of LLM as judge: The pass rate for entities is around 91.33\%, and the pass rate for relation networks is approximately 94.51\%.}
    \label{fig:llmasjudge_show}
\end{figure}


In evaluating the RAKG model, we employed the DeepEval platform \cite{Ip_deepeval_2025} as our assessment tool. It is evident from the results that the output distrust rate of the RAKG model consistently remains at an extremely low level. This stable performance powerfully demonstrates the reliability and consistency of the RAKG model in generating knowledge graphs. Even in the rare instances where hallucinations from the LLM lead to the creation of a small number of false nodes or relationships, we can effectively identify and eliminate these potential errors through the "LLM as judge" mechanism. This process is clearly and intuitively illustrated in Figure \ref{fig:llmasjudge_show}.

\begin{table}[htbp]
\centering
\caption{Merged Matrix with Standard KG Reference}
\label{tab:merged_en}
\resizebox{\columnwidth}{!}{%
\begin{tabular}{l*{2}{>{\centering\arraybackslash}m{3cm}}}
\toprule
\multirow{2}{*}{Target} & \multicolumn{2}{c}{ideal KG as Reference} \\
\cmidrule(lr){2-3}
 & EC & RNS \\
\midrule
KGGen        & 0.6020 $\pm$ 0.1754 & 0.6321 $\pm$ 0.0818 \\
GraphRAG     & 0.6438 $\pm$ 0.1558 & 0.7278 $\pm$ 0.0752 \\
RAKG         & \pmb{0.8752 $\pm$ 0.1047} & \pmb{0.7998 $\pm$ 0.0912} \\ 
\bottomrule
\end{tabular}
}
\end{table}

We conducted an in-depth model performance analysis from two key dimensions: Entity Coverage (EC) and Relation Similarity (RNS). According to the data in Table \ref{tab:merged_en}, the RAKG model achieved an entity coverage of 0.8752 (standard deviation: 0.1047) and a relation similarity of 0.7998 (standard deviation: 0.0912). Both metrics significantly outperformed other models. GraphRAG achieved an entity coverage of 0.6438 and a relation similarity of 0.7278, while KGGen achieved an entity coverage of 0.6020 and a relation similarity of 0.6321. RAKG demonstrated high consistency with the standard knowledge graph across both dimensions, significantly outperforming GraphRAG and KGGen. Further analysis revealed that RAKG achieved the highest entity coverage and exhibited a minor standard deviation, indicating more excellent stability in its results. In terms of relation similarity, RAKG maintained a significant leading advantage, demonstrating its ability to capture relationships between entities more accurately during knowledge graph construction. These results indicate that RAKG has substantial benefits in both the completeness and accuracy of knowledge graph construction, making it better suited for high-quality knowledge graph requirements.

\section{Case Study}


\begin{figure*}[h!]
    \centering
    \includegraphics[width=0.95\textwidth]{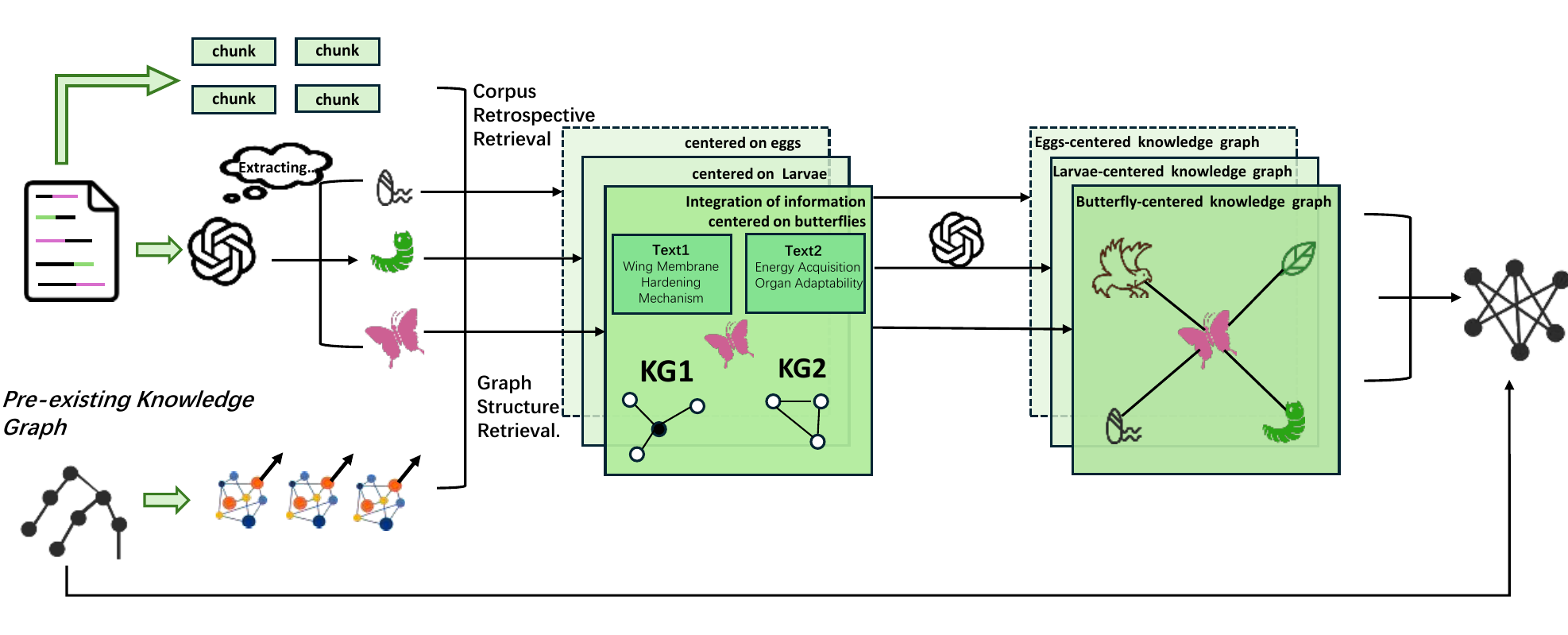} 
    \caption{Case Study. Different colors in the document represent different entities involved. The document is segmented and generates pre-entities such as Butterfly, Caterpillar, and Butterfly Egg. Taking the Butterfly as an example, the texts are retrieved from the text corpus, and related nodes are retrieved from the pre-existing knowledge graph. After integrating the information, the LLM extracts a knowledge graph centered on the butterfly. Finally, these knowledge graphs are integrated.}
    \label{fig:case_study}
\end{figure*}
In the case study, we used an article titled "The Life Cycle of a Butterfly" as the application scenario to compare the performance of the RAKG framework with the baseline models, as shown in Figure \ref{fig:case_study}.
RAKG's named entity recognition module detected 23 core entities in the article, with "Butterfly Egg," "Caterpillar," and "Adult Butterfly" being central. These entities have dense text blocks in the article, indicating key concepts. Focusing on "Adult Butterfly," we retrieved professional text chunks describing five features. We also obtained a sub-graph related to "Adult Butterfly" from the original KG via graph-structured retrieval.
After NER, Corpus Retrospective Retrieval, and Graph Structure Retrieval, we integrated each entity's text blocks with their sub-graphs. These integrated data were fed into a LLM to construct relationship networks. The LLM analyzed the text blocks and sub-graphs to generate entity-specific relationship networks, forming complete sub-graphs.
By integrating all sub-graphs, we built a systematic, structured knowledge graph that clearly shows the article's core concepts and their relationships.

To further evaluate the performance of the RAKG framework, we compared the knowledge graph it generated with the results of baseline models. The results showed that the knowledge graph constructed by RAKG has higher EC and RMS, making it the most similar to the ideal knowledge graph. For example, focusing on the relationship network of the entity "Adult Butterfly", RAKG retrieved the passage "Adult butterflies feed on nectar from flowers using their long, tubular mouthparts called proboscis. They play a crucial role in pollination by transferring pollen from one flower to another, helping plants reproduce." From this, it was concluded that adult butterflies contribute to pollination, resulting in the triple "Adult butterfly"-"contributes to"-"POLLINATION." This demonstrates that RAKG can more comprehensively capture the complex relationships between various entities involved in the different life stages of butterflies.

\section{conclusion}
 In this paper, we propose a novel document-level knowledge graph construction framework named RAKG, which can directly transform document corpora into knowledge graphs. RAKG employs a progressive knowledge extraction method based on pre-entities to integrate information. This approach effectively reduces the complexity of entity disambiguation, circumvents the long-distance forgetting issues of LLMs, and achieves near-perfect performance regarding topological structure coverage and relationship network alignment. The superior performance of RAKG compared to existing state-of-the-art methods demonstrates the effectiveness of our proposed framework.

\bibliographystyle{IEEEtran}
\bibliography{egbib}

\end{document}